\documentclass{jpconf}

\usepackage{graphicx,amsmath,amssymb}

\begin{document}

\title{Critical and multicritical behavior of the $\pm J$ Ising model in two and three dimensions}

\author{M Hasenbusch$^{1}$, F Parisen Toldin$^{2,3}$, A Pelissetto$^{4}$, E Vicari$^{5}$}

\address{$^{1}$ Institut f\"ur Theoretische Physik, Universit\"at Leipzig, D-04009 Leipzig, Germany}
\address{$^{2}$ Max-Planck-Institut f\"ur Metallforschung,
D-70569 Stuttgart, Germany}
\address{$^{3}$ Institut f\"ur Theoretische und Angewandte Physik,
Universit\"at Stuttgart, D-70569 Stuttgart, Germany}
\address{$^{4}$ Dipartimento di Fisica dell'Universit\`a di Roma ``La Sapienza'' and INFN, I-00185 Roma, Italy}
\address{$^{5}$ Dipartimento di Fisica dell'Universit\`a di Pisa and INFN,
I-56127 Pisa, Italy}

\ead{parisen@mf.mpg.de}

\begin{abstract}
We report our Monte Carlo results on the critical and multicritical behavior 
of the $\pm J$ Ising model [with a random-exchange probability 
$P(J_{xy}) = p \delta(J_{xy} - J) + (1-p) \delta(J_{xy} + J)$], 
in two and three dimensions. 
We study the transition line between the paramagnetic and
ferromagnetic phase, which extends from $p=1$ to a multicritical (Nishimori)
point.
By a finite-size scaling analysis, we provide strong numerical evidence that 
in three dimensions the critical behavior along this
line belongs to the same universality 
class as that of the critical transition 
in the randomly dilute Ising model.
In two dimensions we confirm that the critical behavior is controlled by the 
pure Ising fixed point and that disorder is marginally irrelevant, 
giving rise to universal logarithmic corrections.
In both two and three dimensions, we also determine the location of the 
multicritical Nishimori point, as well as the renormalization-group dimensions 
of the operators that control the renormalization-group flow close to it.
\end{abstract}

\section{Introduction}
\begin{figure}
\begin{center}
\includegraphics[width=0.93\linewidth,keepaspectratio]{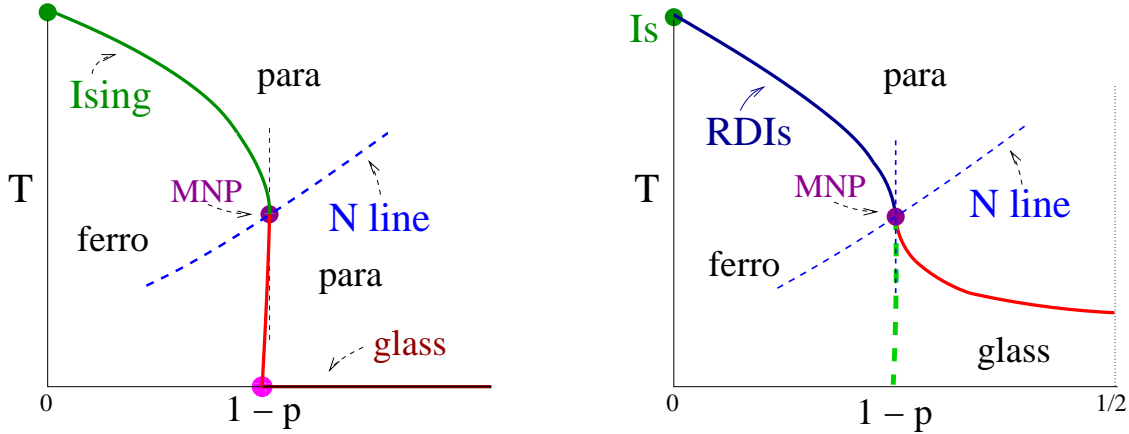}
\end{center}
\caption{Phase diagram of the $\pm J$ Ising model in the $T$-$p$ plane, in dimensions $d=2$ (left) and $d=3$ (right).}
\label{phase}
\end{figure}

The $\pm J$ Ising model is a simplified model \cite{EA-75} for disordered 
uniaxial magnetic 
systems showing glassy behavior in some region of their phase diagram and
represents an interesting
theoretical laboratory to study the effects of quenched disorder and
frustration. It is defined by the lattice Hamiltonian
\begin{equation}
{\cal H} = - \sum_{\langle xy \rangle} J_{xy} \sigma_x \sigma_y,
\label{lH}
\end{equation}
where $\sigma_x=\pm 1$, the sum is over the nearest-neighbor sites of a simple
cubic lattice, and the exchange interactions $J_{xy}$ are uncorrelated
quenched random variables, taking values $\pm J$ with probability distribution
\begin{equation}
P(J_{xy}) = p \delta(J_{xy} - J) + (1-p) \delta(J_{xy} + J). 
\label{probdis}
\end{equation}
In the following we set $J=1$ without loss of generality.
For $p=1$, we recover the standard Ising model, while for $p=1/2$, 
we obtain the usual bimodal Ising spin-glass model.
The phase diagram in two and three dimensions is sketched in 
Figure \ref{phase}; it is symmetric for $p\rightarrow 1-p$ and thus here and 
in the following we only consider the case $1-p<1/2$, i.e., $p > 1/2$.

In both $d=2$ and $d=3$, the model exhibits a paramagnetic-ferromagnetic (PF)
phase transition in the region of low frustration. The PF transition line
starts at the Ising point $X_\text{Is}=(T=T_\text{Is}, p=1)$, 
where $T_\text{Is}$ is the critical temperature of the Ising model,
and extends up to the multicritical Nishimori point (MNP) at 
$X_\text{MNP} = (T^*,p^*)$. 
Along this line, the critical behavior is analogous to that observed in 
randomly dilute Ising (RDI) models.

In two dimensions, renormalization-group (RG) and conformal field theory predict \cite{S-84,S-87,L-90,S-94}
that disorder is marginally irrelevant, thus the critical behavior is controlled
by the pure Ising fixed point, with universal logarithmic corrections \cite{low2d}.
In three dimensions disorder is relevant and the model shows a second-order 
phase transition in the three-dimensional (3D) RDI universality 
class \cite{low3d}, which describes transitions in generic diluted Ising systems with ferromagnetic exchange
interactions \cite{diluted}.

As argued in
\cite{GHDB-85,LH-88,LH-89}, the MNP is
located
along the so-called
Nishimori line ($N$ line)~\cite{Nishimori-81,Nbook} defined by the equation
\begin{equation}
{\rm tanh} \,\beta = 2p-1 ,
\label{nline}
\end{equation}
where $\beta\equiv 1/T$.
At the MNP the transition line is predicted to be parallel
to the $T$ axis \cite{LH-89}. Then, it reaches the $T=0$ axis at $X_c=(0,p_c)$.
It has been proved \cite{Nishimori-81} that
ferromagnetism can only exist in the region $p\ge p^*$; thus,
$p_c$ must satisfy the inequality $p_c\ge p^*$. 
Recent numerical works indicate that $p_c$ is strictly larger than $p^*$,
though deviations are quite small, both in two \cite{nishimori2d} and three \cite{nishimori3d} dimensions, as well as in related models \cite{WHP-03,ONNB-08}.

The critical behavior for $p < p^*$ depends on the dimension. 
The 3D $\pm J$ model exhibits a paramagnetic-glassy (PG) 
transition line, which extends from the MNP up to $p=1/2$. 
On this transition line, the critical behavior is independent on $p$ and 
belongs to the Ising spin-glass universality class \cite{HPV-08}.
In contrast, in two dimensions there is no evidence of a 
finite-temperature glassy phase. 
Glassy behavior is only expected for $T=0$ and $p<p_c$: 
the glassy phase at $T=0$ is unstable with respect to thermal fluctuations.

\section{Finite-size scaling}
\subsection{Paramagnetic-ferromagnetic transition line}
\label{fss.lowf}
We present here the strategy used to analyze the critical behavior on the 
PF transition line in the 3D $\pm J$ model. The 
two-dimensional (2D) case is analyzed in \cite{low2d}.

According to the renormalization group (RG), 
in the case of periodic boundary conditions and for $L\to
\infty$, where $L$ is the lattice size, a generic RG invariant quantity $R$ at
the critical temperature $1/\beta_c$ behaves as
\begin{equation}
R(L,\beta=\beta_c) = R^* \left( 1 + c_{11} L^{-\omega} + 
c_{12} L^{-2\omega} + \cdots + 
c_{21} L^{-\omega_2} + \cdots\right),
\label{Rscal}
\end{equation}
where $R^*$ is the universal infinite-volume limit and $\omega$ and $\omega_2$
are the leading and next-to-leading correction-to-scaling exponents.  
In 3D RDI systems scaling corrections play an important role, 
since $\omega$ is quite small; indeed, we have $\omega=0.29(2)$ and
$\omega_2=0.82(8)$ \cite{HPV-07,diluted}.
Instead of computing the various quantities at fixed Hamiltonian parameters,
we keep a RG invariant quantity $R$ fixed at a given value
$R_{f}$ \cite{Has-99}. This means that, for each $L$, we determine the
pseudocritical inverse temperature $\beta_f(L)$ such that
$R(\beta=\beta_f(L),L) = R_{f}$.
All interesting thermodynamic quantities are then computed at $\beta = \beta_f(L)$.
The pseudocritical inverse temperature $\beta_f(L)$ converges to
$\beta_c$ as $L\to \infty$.  The value $R_{f}$ can be specified at will, as
long as $R_f$ is taken between the high- and low-temperature fixed-point
values of $R$.  The choice $R_{f} = R^*$
(where $R^*$ is the critical-point value)
improves the convergence of $\beta_f$ to $\beta_c$ for $L\to\infty$;
indeed $\beta_f-\beta_c=O(L^{-1/\nu})$ for generic values of $R_f$, while
$\beta_f-\beta_c=O(L^{-1/\nu-\omega})$ for $R_f=R^*$.

We can then consider any other RG invariant quantity $R_\alpha$ 
at fixed $R=R_f$, i.e., $\bar{R}_\alpha(L) = R_\alpha(L,\beta_f(L))$.
For $L\to \infty$, one can show that \cite{diluted}
\begin{align}
\label{Rscalbar}
{\bar R}_\alpha(L) &= {\bar R}^* \left( 1 + b_{11} L^{-\omega} + b_{12} L^{-2\omega} + \cdots + b_{21} L^{-\omega_2} + \cdots \right),\\
\label{Rprimeexp}
{\bar R}_\alpha'(L) &= a_0 L^{1/\nu} \left( 1 + a_{11} L^{-\omega} + a_{12} L^{-2\omega} + \cdots + a_{21} L^{-\omega_2} + \cdots\right),\\
\label{chiscal}
\bar{\chi}(L) \equiv  \chi(L,\beta &= \beta_f(L)) = d_0 L^{2-\eta} \left(  1 + d_{11} L^{-\omega} + d_{12} L^{-2\omega} + \cdots + d_{21} L^{-\omega_2} + \cdots\right) + d_b,
\end{align}
where ${\bar R}_\alpha'$ is the derivative of ${\bar R}_\alpha$ with respect to $\beta$ and $\bar{\chi}$ is the susceptibility at fixed $R=R_f$.
More details on this method can be found in \cite{Has-99,diluted}.

\subsection{Multicritical Nishimori point} \label{fss.nishimori}

In the absence of external fields, the critical behavior at the 
MNP is characterized by two relevant RG operators.
The singular part of the disorder-averaged free energy in a volume $L^d$ 
can be written as
\begin{equation}
F_{\rm sing}(T,p,L) = L^{-d} f(u_1 L^{y_1}, u_2 L^{y_2}, 
\{u_i L^{y_i}\}),\quad i\ge 3,
\label{freeen}
\end{equation} 
where $y_1>y_2>0$, $y_i<0$ for $i\ge 3$, $u_i$ are the corresponding scaling
fields, $u_1 = u_2 = 0$ at the MNP.
The scaling fields $u_i$ are analytic functions of the model parameters $T$
and $p$.  Using symmetry arguments, \cite{LH-88,LH-89} showed that
the scaling axis corresponding to $u_2=0$ is along the $N$ line, so that
$u_2 =  {\rm tanh} \beta-2p+1$.
As for the scaling axis $u_1 = 0$, $\epsilon\equiv 6-d$ expansion calculations
predict it \cite{LH-89} to be parallel to the $T$ axis.  The extension of this
result to lower dimensions suggests $u_1=p-p^*$.
Note that, if this conjecture holds, only the scaling field $u_2$ depends on
the temperature $T$.

These results give rise to the following predictions for the FSS behavior
around $T^*$, $p^*$.
Along the $N$ line,
an RG invariant quantity $R$ has the following behavior for $L\rightarrow\infty$:
\begin{equation}
R_N = R^* + b_{11} u_1 L^{y_1} + \ldots,
\label{RGinvsca}
\end{equation}
where the subscript $N$ indicates that $R$ is restricted to the $N$ line
and we have neglected scaling corrections.
Its derivative $R'$ with respect to $\beta$ behaves as
\begin{equation}
R' = b_{11} u'_1 L^{y_1} + b_{21} u'_2 L^{y_2} + \ldots = b_{21} u'_2 L^{y_2} + \ldots,
\label{RGinvdsca1}
\end{equation}
where the last equivalence holds only if $u_1$ does not depend on the temperature.
This result gives us a method to verify the conjecture of
\cite{LH-89}: once $y_1$ has been determined from the scaling
behavior of a RG invariant quantity $R$ close to the MNP, it is enough to
check the scaling behavior of $R'$. If $R'$ scales as $L^\zeta$ with
$\zeta<y_1$, the conjecture is confirmed and $y_2 = \zeta$.
Finally, along the $N$ line the susceptibility $\chi$ is expected to behave as
\begin{equation}
\chi_N = d_0 L^{2-\eta}\left( 1 + d_1 u_1 L^{y_1} + \cdots\right).
\label{RGinvchi}
\end{equation}

\section{Results}
\subsection{The 3D $\pm J$ model at the paramagnetic-to-ferromagnetic transition line}
In \cite{low3d} we perform MC simulations of Hamiltonian (\ref{lH}) at $d=3$
close to the PF transition line.
Using the method outlined in Sec.~\ref{fss.lowf}, we fit the large-$L$ limit of various RG-invariant quantities at fixed $R_\xi\equiv \xi/L =0.5943$, where $\xi$ is the correlation length; this value is very close to the fixed-point value $R_\xi^*=0.5944(7)$ \cite{diluted} at $\beta_c$.
Our FSS analysis provides strong evidence
that the critical behavior of the  3D $\pm J$ Ising model along the PF 
line belongs to the 3D  RDI universality class.
Indeed, all the RG-invariant quantities considered are in agreement with the results reported in \cite{diluted}.
Moreover, we find $\nu=0.682(3)$ and
$\eta=0.036(2)$, in good
agreement with the presently most accurate
estimates \cite{diluted} $\nu=0.683(2)$ and $\eta=0.036(1)$ for the 3D RDI universality class.

We also note that the random-exchange interaction in the $\pm J$ Ising model
gives rise to frustration, while the RDI universality class describes
transitions in generic diluted Ising systems with ferromagnetic exchange
interactions.  Therefore, our results imply that frustration is irrelevant 
along the PF transition line.
Moreover, the observed scaling corrections are consistent with the RDI 
correction-to-scaling exponents $\omega=0.29(2)$ and
$\omega_2=0.82(8)$ \cite{HPV-07,diluted}. 
Thus, frustration does not introduce new irrelevant perturbations
with RG dimension $|y_f|\lesssim 1$.

\subsection{Multicritical Nishimori point}
In \cite{nishimori2d,nishimori3d}, we perform MC simulations of 
Hamiltonian (\ref{lH}) at $d=2$, $3$, along the $N$ line defined by 
(\ref{nline}).
In both cases, the position of the MNP as well as the RG dimension $y_1$ 
of the leading relevant operator $u_1$ are determined by 
fitting several RG-invariant quantities $R$ to (\ref{RGinvsca}) (in the 2D case
we also consider scaling corrections). We verify the conjecture that 
$u_1$ does not depend on $T$ and compute the RG dimension associated with $u_1$ and $u_2$.

For the 2D model, the MNP is located at $p^*=0.89081(7)$, $T^*=0.9528(4)$, 
the RG dimensions of the operators that control the multicritical behavior are 
$y_1=0.655(15)$ and $y_2 = 0.250(2)$, and the susceptibility exponent is 
$\eta = 0.180(5)$.
In three dimensions, 
we find $p^*=0.76820(4)$, $T^*=1.6692(3)$, $y_1 = 1.02(5)$, $y_2 = 0.61(2)$, 
$\eta = -0.114(3)$.


\section*{References}

\begin{thebibliography}{99}

\bibitem{EA-75}
Edwards S F, Anderson P W 1975
% Theory of spin glasses
{\it J.~Phys.~F: Met.~Phys.} {\bf 5} 965
%%CITATION = JPFMA,5,965;%%

\bibitem{S-84}
Shalaev B N 1984
{\it Sov.~Phys.~Solid State} {\bf 26} 1811

\bibitem{S-87}
Shankar R 1987
% Exact critical behavior of a random bond two-dimensional Ising model
{\it Phys.~Rev.~Lett.} {\bf 58} 2466
%%CITATION = PRLTA,58,2466;%%
\nonum
% Erratum
Shankar R 1987
% Exact critical behavior of a random bond two-dimensional Ising model
{\it Phys.~Rev.~Lett.} {\bf 59} 380
%%CITATION = PRLTA,59,380;%%
\nonum
Ludwig A W W 1988
% Comment on "Exact Critical Behavior of a Random-Bond Two-Dimensional Ising Model"
{\it Phys.~Rev.~Lett.} {\bf 61} 2388
%%CITATION = PRLTA,61,2388;%%
\nonum
Ceccatto H A, Naon C 1988
% Comment on "Exact Critical Behavior of a Random-Bond Two-Dimensional Ising Model"
{\it Phys.~Rev.~Lett.} {\bf 61} 2389
%%CITATION = PRLTA,61,2389;%%
\nonum
Ludwig A W W 1989
% Comment on "Exact Critical Behavior of a Random-Bond Two-Dimensional Ising Model"
{\it Phys.~Rev.~Lett.} {\bf 62} 980
%%CITATION = PRLTA,61,2388;%%

\bibitem{L-90}
Ludwig A W W 1990
% Infinite hierarchies of exponents in a diluted ferromagnet and their interpretation
{\it Nucl.~Phys.} B {\bf 330}
%%CITATION = NUPHA,B330,639;%%

\bibitem{S-94}
Shalaev 1994
% Critical behavior of the two-dimensional Ising model with random bonds
{\it Phys.~Rep.} {\bf 237} 129
%%CITATION = PRPLC,237,129;%%

\bibitem{low2d}
Hasenbusch M, Parisen Toldin F, Pelissetto A, Vicari E 2008
% Universal dependence on disorder of 2D randomly diluted and random-bond +-J Ising models
{\it Phys.~Rev.} E {\bf 78} 011110
%({\it Preprint} arXiv:0804.2788)
%%CITATION = ARXIV:0804.2788;%%

\bibitem{low3d}
Hasenbusch M, Parisen Toldin F, Pelissetto A, Vicari E 2007
% The 3D +-J Ising model at the ferromagnetic transition line
{\it Phys.~Rev.} B {\bf 76} 094402
%({\it Preprint} arXiv:0704.0427)
%%CITATION = PHRVA,B76,094402;%%

\bibitem{diluted}
Hasenbusch M, Parisen Toldin F, Pelissetto A, Vicari E 2007
% Universality class of 3D site-diluted and bond-diluted Ising systems
{\it J.~Stat.~Mech.: Theory Expt.} P02016
%({\it Preprint} arXiv:cond-mat/0611707)
%%CITATION = JSTAT,0702,P016;%%

\bibitem{GHDB-85}
Georges A, Hansel D, Le Doussal P, and Bouchaud J 1985
% Exact properties of spin glasses. II. Nishimori's line : new results and physical implications
{\it J.~Phys.~(Paris)} {\bf 46} 1827
%%CITATION = JOPQA,46,1827;%%

\bibitem{LH-88}
Le Doussal P and Harris A B 1988
% Location of the Ising Spin-Glass Multicritical Point on Nishimori's Line
{\it Phys.~Rev.~Lett.} {\bf 61} 625
%%CITATION = PRLTA,61,625;%%

\bibitem{LH-89}
Le Doussal P and Harris A B 1989
% e expansion for the Nishimori multicritical point of spin glasses
{\it Phys.~Rev.~B} {\bf 40} 9249
%%CITATION = PHRVA,B40,9249;%%

\bibitem{Nishimori-81}
Nishimori H 1981
% Internal Energy, Specific Heat and Correlation Function of the Bond-Random Ising Model
{\it Prog.~Theor.~Phys.} {\bf 66} 1169
%%CITATION = PTPKA,66,1169;%%

\bibitem{Nbook}
Nishimori H 2001
{\it Statistical Physics of Spin Glasses and Information Processing: An Introduction} (Oxford: Oxford University Press)

\bibitem{nishimori2d}
Hasenbusch M, Parisen Toldin F, Pelissetto A, Vicari E 2008
% Multicritical Nishimori point in the phase diagram of the +- J Ising model on a square lattice
{\it Phys.~Rev.} E {\bf 77} 051115
%({\it Preprint} arXiv:0803.0444)
%%CITATION = PHRVA,E77,051115;%%

\bibitem{nishimori3d}
Hasenbusch M, Parisen Toldin F, Pelissetto A, Vicari E 2007
% Magnetic-glassy multicritical behavior of the three-dimensional +- J Ising model
{\it Phys.~Rev.} B {\bf 76} 184202
%({\it Preprint} arXiv:0707.2866)
%%CITATION = PHRVA,B76,184202;%%

\bibitem{WHP-03}
Wang C, Harrington J, Preskill J 2003
% Confinement-Higgs transition in a disordered gauge theory and the accuracy threshold for quantum memory
{\it Ann.~Phys.} {\bf 303} 31
%%CITATION = APNYA,303,31;%%

\bibitem{ONNB-08}
Ohzeki M, Nishimori H, Nihat Berker A 2008
%Multicritical points for the spin glass models on hierarchical lattices
{\it Phys.~Rev.} E {\bf 77} 061116
%({\it Preprint} arXiv:0802.2760)
%%CITATION = PHRVA,E77,061116;%%

\bibitem{HPV-08}
Hasenbusch M, Pelissetto A, Vicari E 2008
% The critical behavior of 3D Ising glass models: universality and scaling corrections
{\it J.~Stat.~Mech.: Theory Expt.} L02001
%({\it Preprint} arXiv:0710.1980)
%%CITATION = JSTAT,0702,L001;%%

\bibitem{HPV-07}
Hasenbusch M, Pelissetto A, Vicari E 2007
% Relaxational dynamics in 3D randomly diluted Ising models
{\it J.~Stat.~Mech.: Theory Expt.} P11009
%({\it Preprint} arXiv:0709.4179)
%%CITATION = JSTAT,0711,P009;%%

\bibitem{Has-99}
Hasenbusch M 1999
% A Monte Carlo study of leading order scaling corrections of phi4 theory on a three-dimensional lattice
{\it J.~Phys.~A: Math. Gen.}  {\bf 32} 4851
%({\it Preprint} arXiv:hep-lat/9902026)
%%CITATION = JPAGB,A32,4851;%%

\end{thebibliography}
\end{document}